\documentclass{elsart}

\usepackage{graphicx}
\usepackage{calrsfs}
\usepackage{amsmath}
\usepackage{amssymb}
\usepackage{citesort}
\usepackage{natbib}

\newcommand{\Log}{\text{Log}\,}
\newcommand{\eps}{\varepsilon}
\newcommand{\epszer}{\eps\rightarrow 0}
\newcommand{\ueps}{u_{\eps}}
\newcommand{\bfna}{\boldsymbol{\nabla}}
\newcommand{\demi} {\frac{1}{2}}
\newcommand{\bfe} {\boldsymbol{e}}
\newcommand{\bfx} {\boldsymbol{x}}
\newcommand{\bfze} {\boldsymbol{0}}
\newcommand{\dip} {\!:\!}
\newcommand{\derp}[2]{\frac{\partial #1}{\partial #2}}
\newcommand{\der}[2]{\frac{\text{d} #1}{\text{d} #2}}
\newcommand{\OOr}{\Omega_{\eps}}
\newcommand{\OO}{\Omega}
\newcommand{\dO}{\partial\Omega}
\newcommand{\ds}{\;\text{d}s}
\newcommand{\dV}{\;\text{d}V}
\newcommand{\iOr}{\int_{\OOr}}
\newcommand{\idO}{\int_{\dO}}
\newcommand{\Br} {B_{\eps}}

\begin{document}

\begin{frontmatter}

\title{Discussion of ``Second order topological sensitivity analysis'' by J. Rocha de Faria et al.\thanksref{foot}}
\thanks[foot]{\emph{International Journal of Solids and Structures, \textbf{44}:4958-4977} (2007).}
\author{Marc Bonnet\corauthref{cor}}
\corauth[cor]{Corresponding author}
\ead{bonnet@lms.polytechnique.fr}
\address{Solid Mechanics Laboratory (UMR CNRS 7649), Ecole
  Polytechnique,\\F-91128 Palaiseau cedex, France}

\begin{abstract}
The aim of this discussion is to expose incorrect results in a previous IJSS article.
\end{abstract}

\end{frontmatter}

\paragraph*{Preliminaries.} The article by \cite{feijoo:07} under discussion is concerned with the evaluation of the perturbation undergone by the potential energy of a domain $\OO$ (in a 2-D, scalar Laplace equation setting) when a disk $\Br$ of small radius $\eps$ centered at a given location $\hat{\bfx}\in\OO$ is removed from $\OO$, assuming either Neumann or Dirichlet conditions on the boundary of the small `hole' thus created. In each case, the potential energy $\psi(\OOr)$ of the punctured domain $\OOr=\OO\setminus\Br$ is expanded about $\eps=0$ so that the first two terms of the perturbation are given. The first (leading) term is the well-documented topological derivative of $\psi$. The article under discussion places, logically, its main focus on the next term of the expansion. However, it contains incorrrect results, as shown in this discussion. In what follows, equations referenced with Arabic numbers refer to those of the article under discussion.

\paragraph*{Topological expansion: Neumann condition on the hole.} In the main result proposed by \cite{feijoo:07} for this case, namely expression~(37) for the topological expansion of the potential energy, the first term (whose order is $O(\eps^{2})$) is correct but the second (whose order is $O(\eps^{4})$) is not as it lacks a contribution of the same order related to the external boundary \citep[see][for a similar study in 3-D linear acoustics]{B-2006-6}.

This error can be explained as follows. Equation~(37) is based on an expansion of
\[
  \der{}{\eps}\psi(\OOr)
 = -\demi\int_{\partial\Br} (\bfna\ueps.\bfe_{\theta})^{2} \ds \tag{i}
\]
up to order $O(\eps^{3})$ (where $(\bfe_{r},\bfe_{\theta})$ are the unit vectors associated with polar coordinates $(r,\theta)$ originating at the center of $\Br$). Since $\ds=\eps\text{d}\theta$ on $\partial\Br$, this task requires expanding $(\bfna\ueps(\bfx).\bfe_{\theta})^{2}$ to order $O(\eps^{2})$ for $\bfx\in\partial\Br$. The latter operation is carried out in~\cite{feijoo:07} by evaluating $\bfna\ueps(\bfx)$ from the $O(\eps^{2})$ expansion~(23) of $\ueps$. However, expansion~(23) evaluated on $\partial\Br$ gives
\[
 \bfna\ueps(\bfx).\bfe_{\theta}
 = 2\bfna u(\hat{\bfx}).\bfe_{\theta}
 + 2\eps\bfna\bfna u(\hat{\bfx})\dip(\bfe_{r}\otimes\bfe_{\theta}) + O(\eps^{2})
\qquad (\bfx\in\partial\Br),
\]
and is therefore not suitable for expanding $(\bfna\ueps.\bfe_{\theta})^{2}$ to order $O(\eps^{2})$ as it lacks the necessary $O(\eps^{2})$ contribution to $\bfna\ueps.\bfe_{\theta}$. The missing $O(\eps^{2})$ term stems from the $O(\eps^{3})$ contribution to $\ueps$ and is in fact non-local as it is expressed in terms of quantities on $\dO$ rather than higher-order gradients of $u$ at $\hat{\bfx}$.

The incorrectness of result~(37) can be further demonstrated on a simple analytical example.  Consider the 2-D domain $\OOr$ enclosed by two concentric circles of radii $\eps$ and $a$, i.e. $\partial\Br=\{(r,\theta)\,\bigl|\,r=\eps\}$ and $\dO=\{(r,\theta)\,\bigl|\,r=a\}$ in terms of polar coordinates $(r,\theta)$. The solution $\ueps$ of the Laplace equation with boundary conditions 
\[
 u_{,n} = 0 \; (r=\eps), \quad  u_{,n}\equiv q=\cos\theta \; (r=a)
\]
and the corresponding reference solution $u$ when there is no hole are respectively given (up to an arbitrary additive constant) by
\[
  \ueps(r,\theta)
 = \dfrac{a^{2}}{a^{2}-\eps^{2}} \Bigl( r+\dfrac{\eps^{2}}{r} \Bigr)\cos\theta,\qquad
  u(r,\theta)
 = r\cos\theta \tag{ii}
\]
Note that the reference solution $u$ is such that $\bfna u(\hat{\bfx})=\cos\theta\bfe_{r}-\sin\theta\bfe_{\theta}$ and $\bfna\bfna u(\hat{\bfx})=\bfze$. Then, a simple calculation gives
\[
  \psi(\OOr) = \demi\iOr \bfna\ueps.\bfna\ueps \dV - \idO q\ueps \ds
 = -\demi \idO q\ueps \ds
 = -\dfrac{\pi a^{2}}{2} \dfrac{a^{2}+\eps^{2}}{a^{2}-\eps^{2}}
\]
Expanding $\psi(\OOr)$ to order $O(\eps^{4})$ gives
\[
  \psi(\OOr)
 = -\dfrac{\pi a^{2}}{2} - \pi \eps^{2} - \dfrac{\pi}{a^{2}}\eps^{4} + o(\eps^{4})
\tag{iii}
\]
while equation~(37) incorrectly gives the expansion as
\[
  \psi(\OOr)
 = -\dfrac{\pi a^{2}}{2} - \pi \eps^{2} - 0\times \eps^{4} + o(\eps^{4}) \tag{iv}
\]
Note that the error in~(iv) vanishes as $\dO$ is rejected to infinity, i.e. as the influence of the external boundary goes away. This is analogous to secondary reflection effects in small-obstacle approximations for wave problems.

\paragraph*{Topological expansion: Dirichlet condition on the hole.} The topological expansion~(38) is also not correct. Expansion~(38) states that 
\[
  \psi(\OOr) = \psi(\OO) + \pi\Bigl(\dfrac{-1}{\Log\eps}\Bigr)[u(\hat{\bfx})]^{2}
 + \pi\|\bfna u(\hat{\bfx})\|^{2}\eps^{2} + o(\eps^{2}).
 \tag{v}
\]
However, another simple analytical example again allows to show that the second term in~(v), is not correct.  With the domain $\OOr$ defined as before, the solution $\ueps$ of the Laplace equation with boundary conditions 
\[
 u = 0 \; (r=\eps), \qquad  u=A \; (r=a)
\]
and the corresponding reference solution $u$ are respectively given by
\[
  \ueps(r,\theta)
 = A\dfrac{\Log(r/\eps)}{\Log(a/\eps)} ,\qquad u(r,\theta) = A
\]
The potential energy is therefore
\[
 \psi(\OOr) = \demi\iOr \bfna\ueps.\bfna\ueps \dV = \demi\idO \derp{\ueps}{n}\ueps \ds
 = \dfrac{\pi A^{2}}{\Log(a/\eps)}
 = \dfrac{\pi A^{2}}{\Log a-\Log\eps}.
\]
Expanding the above result in powers of $-1/\Log\eps$ yields
\[
  \psi(\OOr) = \pi A^{2} \biggl[ \Bigl(\dfrac{-1}{\Log\eps}\Bigr)
 + \Log a \Bigl(\dfrac{-1}{\Log\eps}\Bigr)^{2} \biggr]
 + o \Bigl(\;\Bigl(\dfrac{-1}{\Log\eps}\Bigr)^{2}\Bigr) \tag{vi}
\]
Expansion~(vi) implies that
\[
   \dfrac{1}{\eps^{2}} \biggl[ \psi(\OOr) - \psi(\OO)
 - \pi\Bigl(\dfrac{-1}{\Log\eps}\Bigr)[u(\hat{\bfx})]^{2} \biggr] \longrightarrow\infty
  \qquad (\epszer)
\]
(noting that $\psi(\OO)=0$ for this example) which directly contradicts expansion~(v), i.e.~(38), except possibly in the special case $a=1$.

\paragraph*{References.} The authors of~\cite{feijoo:07} were apparently not aware of recent references directly related to their work, in particular studies concerned with small-defect asymptotic expansions \citep[e.g.][and works cited therein]{ammari:kang,vog:volkov,volkov} and with the topological derivative for 3-D in the context of scalar and elastic wave propagation \citep{boj:07,B-2005-6,B-2005-17}.

\end{document}